\documentclass[aps,preprint,tightenlines,showpacs]{revtex4}
\usepackage{bm,dcolumn,graphicx}

\begin{document}
\title{
On the feasibility of cooling and trapping metastable alkaline-earth
atoms}

\author{Andrei Derevianko\email{andrei@unr.edu}}
\affiliation { Department of Physics, University of Nevada, Reno,
Nevada 89557-0058}

\date{\today}

\begin{abstract}
Metastability and long-range interactions of Mg, Ca, and Sr in the
lowest-energy metastable $^3P_2$ state are investigated.
 The calculated lifetimes are 38 minutes for Mg$^*$, 118
minutes for Ca$^*$, and 17 minutes for Sr$^*$, supporting
feasibility of cooling and trapping experiments. The
quadrupole-quadrupole long-range interactions of two metastable
atoms are evaluated for various molecular symmetries. Hund's case
(c) $4_g$ potential possesses a large 100-1000 K potential
barrier. Therefore magnetic trap losses can possibly be reduced
using cold metastable atoms in a stretched $M=2$ state.
Calculations were performed in the framework of {\em ab initio}
relativistic configuration interaction method coupled with the
random-phase approximation.
\end{abstract}

\pacs{31.10.+z, 34.20.Cf, 32.10.Dk, 31.15.Ar}

\maketitle

\section{Introduction}

This work is motivated by emerging experiments on cooling and
trapping of alkaline-earth atoms ( see, e.g.,
Ref.~\cite{ZinBinRie00,KatIdoIso99,DinVogHal99}). In particular,
the long-lived lowest-energy $^3\!P_2$ state can serve as an
effective ground state in such experiments. Recently Katori {\em
et al.}~\cite{KatIdoIso00}  cooled and trapped metastable Sr and
proposed a scheme for a continuous atom laser with a possible
application in atom lithography. The purpose of this work is to
evaluate properties of metastable $nsnp\,^3\!P_2$ states for Mg
($n=3$), Ca ($n=4$), and Sr ($n=5$). In particular, we calculated
decay rates of the $nsnp \,^3\!P_2$ states. The resulting
lifetimes are on the order of $10^3-10^4$ s supporting the
feasibility of the experiments.

Ultracold collision properties, including scattering lengths,
are sensitive to  long-range atomic interactions. The dominant
van der Waals interaction of two atoms in their respective $^3P_2$
states is described in terms of the quadrupole moment of the
atomic state. To assist in determining  molecular potentials,
the atomic quadrupole moments of Mg$^{*}$, Ca$^{*}$, and Sr$^{*}$
are also calculated here, and the relevant $C_5$ coefficients for
various molecular symmetries are tabulated. These coefficients are
substantially larger compared to those for metastable
noble-gases~\cite{DoeVreOdB98}. In particular, Hund's case (c)
$4_g$ potential possesses a large 100-1000 K  potential barrier.
Therefore magnetic-trap losses can possibly be reduced using
cold metastable atoms in a stretched $M=2$
state~\cite{KatPrivate}.

\section{ Method}
{\em Ab initio} relativistic valence
configuration-interaction (CI) method coupled with random-phase
approximation (RPA) was employed here. A detailed description of
this method~\cite{Joh00} will be published elsewhere; only a brief
discussion is presented here.  In this method the wave-function is
expanded in terms of two-particle basis functions as
\begin{equation}
\Psi \left( \pi JM\right) = \sum_{k\geq l}
c_{kl}\,\,\Phi_{kl}\left( \,\pi JM\right) \, .
\end{equation}
 Here $J$ is the total
angular momentum with projection $M$, and  $\pi$ is the parity
of the state $\Psi$. The weights $c_{kl}$ and energies are found
by solving the eigen-value problem based on  the {\em no-pair}
Hamiltonian~\cite{BroRav51}. The basis functions are defined in
the subspace of virtual orbitals
\begin{equation}
\Phi _{kl}\left( \,\pi JM\right) =
\eta _{kl}\sum_{m_{k}m_{l}}
C_{j_{k}m_{k},j_{l}m_{l}}^{JM}\,
a_{\{n_{k}\kappa _{k}m_{k}\}}^{\dagger }
a_{\{n_{l}\kappa _{l}m_{l}\}}^{\dagger }|0_{\rm core} \rangle \, ,
\label{Eqn_bas}
\end{equation}
where sets $\{n \kappa m \}$  enumerate  quantum numbers, $
\eta_{kl}^2 = 1 - \frac{1}{2} \delta_{n_k n_l} \delta_{\kappa_k
\kappa_l}$ is a normalization factor, $a^{\dagger}$ are creation
operators, and the quasi-vacuum state $|0_{\rm core} \rangle$
corresponds to a closed-shell core. The one-particle orbitals are
found in the-frozen core ($V^{N-2}$) approximation, i.e. the
Dirac-Hartree-Fock (DHF) equations are solved self-consistently
for core orbitals, and the virtual orbitals are determined using
the resulting DHF potential of the core. The employed set of basis
functions in  Eq.~(\ref{Eqn_bas}) is essentially complete, and one
could interpret the solution of the eigen-value problem as
treating the strong Coulomb repulsion of the two valence electrons
to all orders of perturbation theory.  This valence-CI method
being exact for He-like systems~\cite{JohPlaSap95}, represents an
approximation for the alkaline-earth systems. The
core-polarization effects and the Breit interaction are neglected
here.

Once the wave-functions are determined, the matrix elements of a
one-particle operator $Z=\sum_{ij} z_{ij} a^{\dagger}_i a_j  $ are
computed by forming products $\langle \Psi_F | Z | \Psi_I
\rangle$. The additional RPA approximation~\cite{AmuChe75},
describing shielding of an externally applied field by core
electrons, constitutes a substitution of the ``bare'' one-particle
matrix elements $z_{ij}$ by ``dressed'' matrix elements
$z_{ij}^{\rm RPA}$. Such an approach sums a certain class of
many-body diagrams to all orders of perturbation theory.

The configuration-interaction eigenvalue problem was
solved numerically using the B-spline basis set technique~\cite{JohBluSap88}. The
employed basis set included the partial waves  $s_{1/2}$--$h_{11/2}$.
For each partial wave, 40 positive-energy basis functions
approximated by B-splines represented a complete set. An inclusion
of the 30 lowest-energy basis functions for each partial wave was
found to be sufficient for the goals of this work.

\section{ Lifetime of $^3\!P_2$ states}
\begin{figure}
\centerline{\includegraphics*[scale=1.0]{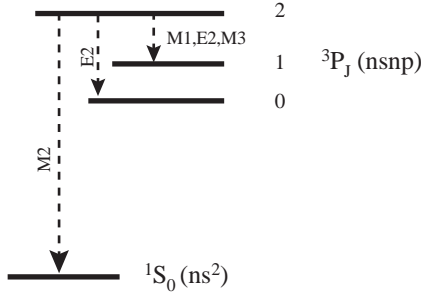}} %
\caption{ Single-photon decay channels from the lowest-energy
$^3\!P_2$ state of alkaline-earth atoms. \label{Fig_Ediag} }
\end{figure}

Single-photon transitions from the metastable $^3\!P_2$ state are
presented in Fig.~\ref{Fig_Ediag}. It will be demonstrated that
for Mg and Ca this level predominantly decays through
magnetic-quadrupole (M2) transition to the ground $^1\!S_0$ state.
For Sr the main decay channel is a magnetic-dipole (M1) transition
to the $J=1$ level of the same $^3\!P$ fine-structure multiplet.
Estimates show that the magnetic-octupole (M3) $^3\!P_2
\rightarrow ^3\!P_1$ transition and all second-order two-photon
decays contribute a negligibly small fraction to the total decay
rate. Further, cold-atom experiments are focused on isotopes
without nuclear spin to avoid negotiating molecular potentials
complicated by hyperfine structure; we will not consider otherwise
important electric-dipole hyperfine-structure induced
decays~\cite{Gar67}.

The data on magnetic-quadrupole transitions (M2) are scarce.
For Mg the $^3\!P_2- ^1\!S_0$ M2 rates were estimated by
Mizushima~\cite{Miz66} and Garstang~\cite{Gar67} more than three
decades ago and recently by J\"{o}nsson and
Fischer~\cite{JonFis97}. There are no published data on M2
transition rates for Ca and Sr.

A general treatment of multipole electromagnetic transitions in
the relativistic framework can be found, for example, in
Ref.~\cite{JohPlaSap95}. The Einstein coefficient $A_{M2}$ for
a magnetic-quadrupole transition $I \rightarrow F$ is given
by\footnote{Unless specified otherwise, atomic units
$\hbar=|e|=m_e=1$ are used throughout the paper.}
\begin{equation}
 A_{M2} = \frac{1}{60} \alpha^7 \omega^5
 \frac{|\langle \Psi_F|| {\mathcal M}^{(2)}|| \Psi_I \rangle|^2 }{ 2J_I +1} \, ,
\end{equation}
where $\omega$ is the photon frequency and $\alpha$ is the
fine-structure constant. The relevant one-particle reduced matrix
element can be expressed as
\begin{eqnarray}
\langle \phi_i || {\mathcal M}^{(2)} || \phi_j \rangle &=&
\frac{2}{3 \, \alpha}
 \langle -\kappa_i || C^{(2)} || \kappa_j \rangle
 (\kappa_i+\kappa_j) \times  \,
\nonumber \\ &&
  \int_0^\infty
 \left\{15 \, \frac{j_2(k r)}{ (k r)^2 }  \right\} r^2
 [ G_i(r) F_j(r) + F_i(r) G_j(r) ] dr \,.   \label{Eq_melM2}
\end{eqnarray}
Here $C^{(2)}$ is the normalized spherical
harmonic~\cite{VarMosKhe88}, $G(F)$ are large (small) radial
components of a wavefunction, $\kappa=(l-j)(2j+1)$,  and photon
wavevector is defined as $k = \omega/c$.
The term in curly brackets approaches unity in the long-wavelength
approximation. Although the full retarded form of the matrix
element was used in the calculations, the long-wavelength limit is
well satisfied for transitions between the lowest-energy valence
states of alkaline-earth atoms. While the matrix element is
largely independent of the transition frequency $\omega$, the rate
$A_{M2}$ depends on it strongly ( $\sim \omega^5$ ) and the
experimental energy intervals were used in the final tabulation.

The calculated CI+RPA magnetic-quadrupole transition rates are
presented in Table~\ref{Tab_M2}. It was found that the
RPA-dressing of matrix elements affects these rates only on the order of 1\%.
The calculated M2 transition rate for Mg$^*$, $4.41 \times
10^{-4}$ 1/s, is in a fair agreement with the multiconfiguration
Dirac-Fock result, $3.983 \times 10^{-4}$ 1/s, by J\"{o}nsson and
Fischer~\cite{JonFis97}. Both rates are as twice as large compared
to the earlier estimate~\cite{Gar67}, $1.6 \times 10^{-4}$
1/s.

Magnetic-dipole $^3\!P_2 \rightarrow ^3\!P_1$ transitions occur
between the levels of the same fine-structure multiplet. The
associated decay rate can be reliably estimated in the
nonrelativistic approximation as
\begin{equation}
 A_{M1} = \frac{1}{6} \alpha^5 \omega^3  \, .
\end{equation}
The M1 rates, presented in Table~\ref{Tab_M2}, were computed using
experimental energy intervals. The determined rate for Mg is 3\%
larger than in calculations~\cite{JonFis97} where theoretical
fine-structure splittings were employed. The discrepancy is
accounted for by a difference between theoretical and experimental
energies.

We also calculated electric-quadrupole (E2) decay rates to the
$^3\!P_{1,0}$ fine-structure levels. The results listed in
Table~\ref{Tab_M2} show that the contribution of these decay
channels is small compared to the M2 and M1 rates.

From the analysis of Table~\ref{Tab_M2} one finds that for Mg and
Ca the $^3\!P_{2}$ state predominantly decays through
magnetic-quadrupole (M2) transition to the ground $^1\!S_0$ state.
For Sr the main decay channel is a magnetic-dipole (M1) transition
to the $J=1$ level of the same $^3\!P$ fine-structure multiplet.
Overall, the calculated lifetimes for Mg$^*$,Ca$^*$, and Sr$^*$
are on the order of $10^3-10^4$ seconds, thus favoring the usage of
these metastable alkaline-earth atoms in cooling and trapping
experiments.

\section{ Quadrupole-quadrupole long-range interactions}
Ultracold collision properties, including scattering lengths,
are sensitive to long-range atomic interactions. Here we
focus on the long-range interactions of two alkaline-earths atoms
in their respective $^3\! P_2$ states.  At large internuclear
separations, $R$, atomic wavefunctions are perturbed by the
axially-symmetric molecular potential, which itself depends on the electronic
coordinates and the internuclear distance. This potential can be
expanded in a sum of interactions of various atomic
multipoles~\cite{DalDav66}. The lowest-order correction to
molecular term energies correlating to two $^3\! P_2$ atomic
states arises from a quadrupole-quadrupole contribution~\cite{Kni38}
\begin{equation}
V_{QQ} = \frac{1}{R^5}
\sum_{\mu =-2}^{2}\frac{4!}{(2-\mu )!(2+\mu )!}
\left(Q_{\mu}\right)_{\rm I}  \left(Q_{-\mu }\right)_{\rm II}\, .
\label{Eqn_QQ}
\end{equation}
Here subscripts I and II denote the subspaces of the electronic coordinates of the two atoms,
and the quadrupole spherical tensor is defined as
\begin{equation}
Q_{\mu }=-|e|\sum_{i}r_{i}^{2} C_{\mu }^{(2)}(\hat{{\mathbf r}}_{i}) \, ,
\label{Eq_Qten}
\end{equation}
where the summation is over atomic electrons,
${\mathbf r}_i$ is the position vector of electron $i$,
and $C_{\mu }^{(2)}(\hat{{\mathbf r}}_{i})$
are normalized spherical harmonics~\cite{VarMosKhe88}.

The quadrupole-quadrupole interactions are parameterized
by the quadrupole moment ${\mathcal Q}$ of the $^3\! P_2$ atomic state,
defined conventionally as
\begin{equation}
 {\mathcal Q} = 2 \, \, \langle ^3\! P_2, M_J=2|\,Q_0\,|^3\! P_2, M_J=2 \rangle \, .
\end{equation}
This quadrupole moment is related to the reduced matrix element of the
tensor, Eq.~(\ref{Eq_Qten}), as
$
{\mathcal Q}(^3\! P_2) = \sqrt{\frac{8}{35}} \,  \langle ^3\! P_2||Q||^3\! P_2 \rangle.
$
The associated one-particle reduced matrix element is given by
\begin{equation}
\langle \phi_i || Q || \phi_j \rangle =
 \langle \kappa_i || C^{(2)} || \kappa_j \rangle
 \,
  \int_0^\infty  r^2
 [ G_i(r) G_j(r) + F_i(r) F_j(r) ] dr \,.   \label{Eq_melQ}
\end{equation}
Using the valence CI-method coupled with the RPA dressing of
matrix elements we calculated the quadrupole moments (see
Table~\ref{Tab_QM}). It was found that the RPA sequence of
diagrams reduces the final result only by 0.1\% for Mg, 0.3\% for
Ca, and 0.5\% for Sr. Due to the diffuse nature of valence states,
these quadrupole moments are significantly larger than those found
for metastable noble-gas atoms~\cite{DoeVreOdB98}, where the hole
in the outer $p_{3/2}$ subshell determines the  quadrupole moment.

As in the case of metastable $^3\! P_2$ noble-gas atoms~\cite{DoeVreOdB98,DerDal00},
the long-range molecular potentials of metastable alkaline-earth
atoms depend on their spatial orientation,
i.e. the interactions are anisotropic.
Altogether, 15 distinct molecular states correlate to the two atomic $^3\! P_2$ states at
large separations. The first order correction to a
molecular term may be represented as
\begin{equation}
U^{(1)}(R) = \frac{C_5 }{R^5}\,.
\end{equation}
The relevant constants $C_5$ for various molecular symmetries are
given in Table~\ref{Tab_C5}, where the states are characterized
using Hund's case (c). These coefficients were obtained by
diagonalizing the quadrupole-quadrupole interaction,
Eq.~(\ref{Eqn_QQ}), in the basis of products of atomic states
\mbox{$| ^3P_2, M \rangle_{\rm I} \otimes | ^3P_2, M' \rangle_{\rm
II}$} for a given symmetry $\Omega$ satisfying $M + M' = \Omega$.

\begin{figure}
\centerline{\includegraphics*[scale=0.75]{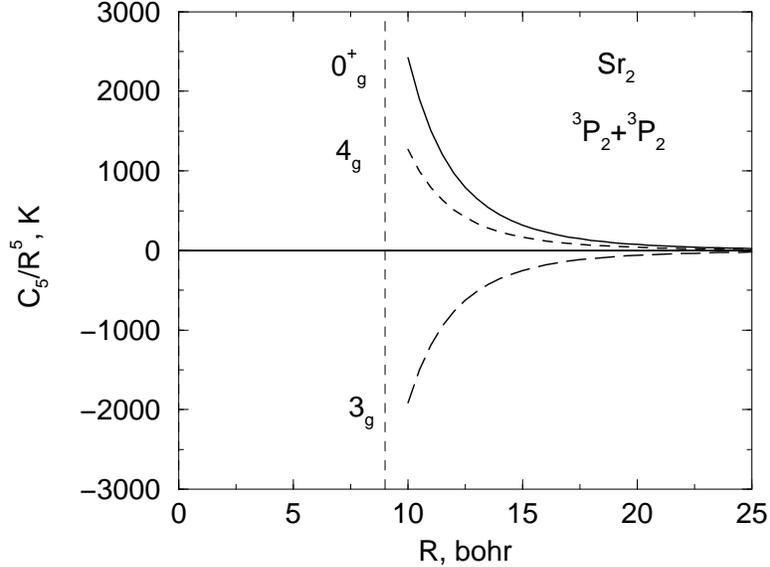}} \caption{ The
most repulsive $0^+_g$ and the most attractive $3_g$ long-range
quadrupole-quadrupole interaction potentials correlating to two
metastable $5s5p\,^3\!P_2$ Sr atoms. The $4_g$ potential is also
shown.  1 K $= 0.69503$ cm$^{-1}$. \label{Fig_QQ} }
\end{figure}

Seven of the resulting quadrupole-quadrupole long-range potentials
are attractive at large distances and eight are repulsive. We
present the most repulsive $0^+_g$ and the most attractive $3_g$
potentials in Fig.~\ref{Fig_QQ}. It is worth discussing the
repulsive $4_g$ potential also shown in Fig.~\ref{Fig_QQ}.
Provided the metastable atoms in a magnetic trap are prepared in a
stretched $M=2$ state, the collisions would occur along this
repulsive $\Omega=4$ potential. Although the medium-range part of
the potential is attractive, the resulting barrier will be on the
order of 100-1000 K high.  Therefore ultracold collisions of the
$^3\!P_2$ alkaline-earth atoms in a stretched $M=2$ state can be
effectively shielded from the losses occurring at small
internuclear distances~\cite{KatPrivate}. This barrier is more
pronounced compared to metastable noble-gas
atoms~\cite{DoeVreOdB98}, where the potential barriers were  a few
nK high. Clearly, calculations of the second-order $C_6$ dispersion
coefficients and intermediate- and short-range parts of the
potentials are needed for a more quantitative description.

\section{Conclusion}
To address the needs of emerging experiments on cooling and
trapping of alkaline-earth atoms, we performed {\em ab initio}
relativistic calculations of lifetimes and quadrupole moments of
metastable Mg, Ca, and Sr. The determined lifetimes are 38 minutes
for Mg$^*$, 118 minutes for Ca$^*$, and 17 minutes for Sr$^*$,
supporting the feasibility of experiments. In
addition, we investigated long-range quadrupole-quadrupole
interactions for molecular potentials correlating to two
metastable atoms. Several resulting potentials possess pronounced
barriers, which could be exploited to minimize trap losses.

Thanks are due to H. Katori, C. Oates, and F. Riehle for
stimulating discussions and V. Davis for careful reading of the manuscript.
The developed numerical code was partially based on programs by Notre
Dame group led by W.R. Johnson. This work was supported in part by the Chemical Sciences,
Geosciences and Biosciences Division of the Office of Basic Energy
Sciences, Office of Science, U.S. Department of Energy.

\begin{table}
\centering \caption{Einstein coefficients $A$ for decays from the
lowest-energy $^3\!P_2$ states in s$^{-1}$. Notation $x[y]$ stands
for $x \times 10^y$. \label{Tab_M2}}
\begin{tabular}{lrrr}
\hline\hline
\multicolumn{1}{c }{Type, final state} %
& \multicolumn{1}{c }{Mg} %
& \multicolumn{1}{c }{Ca} %
& \multicolumn{1}{c }{Sr} %
         \\
\hline M2,$ns^2\,^1\!S_0 $ &
              4.41[-4] &  1.25[-4]& 1.27[-4]      \\
M1, $nsnp\,^3\!P_1 $   &
              9.12[-7] &  1.60[-5]& 8.26[-4]      \\
E2, $nsnp\,^3\!P_1 $   &
              1[-12]   &  3[-10]  &  3[-7]\\
E2, $nsnp\,^3\!P_0 $   &
              3[-12]   &  1[-9]  &  1[-6]\\
\hline
 $A$, total            &
            4.42[-4]   &  1.41[-4]& 9.55[-4]\\
\hline\hline
\end{tabular}
\end{table}

\begin{table}
\centering \caption{ Quadrupole moments ${\mathcal Q}$ of the
lowest-energy metastable $^3\! P_2$ states in a.u. \label{Tab_QM}}
\begin{tabular}{ccc}
\hline\hline
 Mg     & Ca     & Sr  \\
\hline
 8.59   & 13.6   & 16.4\\
\hline\hline
\end{tabular}
\end{table}

\begin{table}
\centering \caption{ $C_5$ coefficients in units of ${\mathcal
Q}^2$ for molecular states ( Hund's case (c)) asymptotically
correlating to two $^{3}\!P_2$ state atoms. The quadrupole moments
${\mathcal Q}$ are listed in Table~\protect\ref{Tab_QM}, and
long-range interaction potentials are parameterized as $U(R)=
C_5/R^5$. \label{Tab_C5}}
\begin{tabular}{ldld}
\hline\hline
\multicolumn{1}{c }{$\Omega$} & \multicolumn{1}{c
}{$C_5/ {\mathcal Q}^2$}& \multicolumn{1}{c }{$\Omega$} &
\multicolumn{1}{c }{$C_5/ {\mathcal Q}^2$} \\
\hline
$0^+_g$ &  2.85329 & $3_g$ & -2.25  \\
$0^-_u$ &  2.42705 & $2_u$ & -1.75  \\
$4_g$   &  1.5     & $2_g$ & -1.625 \\
$0^+_g$   &  1.31989 & $1_g$ & -1.23602\\
$1_u$     &  1.05202 & $0^{-}_u$ & -0.92705\\
$1_g$     &  0.98602 & $1_u$     & -0.80203\\
$2_g$     &  0.75    & $0^+_g$ & -0.42319 \\
$3_u$     &  0.75    &     \\
\hline\hline
\end{tabular}
\end{table}

\end{document}